\documentclass[aps,pra,twocolumn,showpacs,amsmath,amssymb,superscriptaddress,longbibliography,floatfix]{revtex4-2}

\usepackage{graphicx}
\usepackage{dcolumn}
\usepackage{bm}
\usepackage{hyperref}
\hypersetup{colorlinks=true,allcolors=blue}
\usepackage[all]{hypcap}

\usepackage{xcolor}
\usepackage{orcidlink}

\begin{document}

\title{Avoided crossings in spin-boson systems: consequences for adiabatic state preparation}

\author{Joona Marjamäki%
\orcidlink{0009-0009-6680-8057}}
\email{joona.a.marjamaki@jyu.fi}
\affiliation{Department of Physics, Nanoscience Center, P.O. Box 35, 40014 University of Jyväskylä, Finland}
\author{Robert van Leeuwen%
\orcidlink{0000-0002-2499-9125}}
\affiliation{Department of Physics, Nanoscience Center, P.O. Box 35, 40014 University of Jyväskylä, Finland}
\author{Riku Tuovinen%
\orcidlink{0000-0002-7661-1807}}
\email{riku.m.s.tuovinen@jyu.fi}
\affiliation{Department of Physics, Nanoscience Center, P.O. Box 35, 40014 University of Jyväskylä, Finland}

\date{\today}

\begin{abstract}
Avoided crossings in the correlated many-body spectrum play a central role in determining the conditions for adiabatic state preparation in interacting quantum systems. We investigate this connection in finite spin-boson systems by combining exact diagonalization of the many-body spectrum with time-dependent simulations of adiabatic switching protocols. By comparing the adiabatically prepared state with the exact ground state, we determine the switching times required for reliable preparation as a function of the spin-boson coupling strength and bosonic mode structure. We show that regions of small many-body level spacings, associated with avoided crossings, impose the strongest constraints on adiabatic preparation and explain the observed dependence of the required switching times on the interaction strength. Furthermore, while the ground-state energy follows the overall spectral trends, local observables such as Pauli matrix expectation values exhibit additional non-monotonic behavior arising from coherent oscillations in finite systems. These findings can provide concrete benchmarks for both the theoretical treatment of open quantum systems and practical implementations in quantum simulation and qubit control.
\end{abstract}

\maketitle

\section{\label{sec:introduction}Introduction}

The interaction between a quantum system and its environment is a central problem in modern quantum physics~\cite{caldeira_quantum_1983, leggett_dynamics_1987}, with direct implications for the development of quantum technologies and fundamental questions of decoherence and equilibration~\cite{fauseweh_quantum_2024, gatto_quantum_2024}. A paradigmatic model in this context is the spin-boson model, which describes a two-level system coupled to a bosonic environment~\cite{breuer_theory_2002}. This model has played a foundational role in the theory of dissipative quantum systems and serves as a vital benchmark for studying decoherence, quantum phase transitions, and non-equilibrium dynamics~\cite{sun_quantum_2025}. Specifically, it provides a minimal description of solid-state qubits interacting with electromagnetic or vibrational environments, making it highly relevant for quantum computing platforms~\cite{blais_circuit_2021, schlimgen_quantum_2021}.

In most practical applications, the bosonic environment is modeled as a continuum bath and treated using approximate methods such as weak-coupling expansions, Markovian master equations, or perturbative techniques~\cite{breuer_theory_2002, xu_colloquium_2026}. While these approaches are computationally efficient, they rely on assumptions that frequently break down in regimes of strong coupling, highly structured environments, or finite-size baths~\cite{chin_exact_2010, devega_dynamics_2017, vadimov_validity_2021}. In such cases, the separation between system and environment becomes ambiguous, and the notion of equilibration requires careful consideration~\cite{eisert_quantum_2015, gogolin_equilibration_2016}. Moreover, the commonly employed partitioned initial condition, where the system and bath are initially uncoupled, can introduce non-physical transients that obscure the intrinsic physics of the coupled steady state, motivating approaches that prepare correlated equilibrium states prior to the subsequent time evolution.

An alternative perspective involves considering finite bosonic environments that can be treated exactly, which allows isolating intrinsic quantum dynamics from bath-induced approximations~\cite{steib_nonlinear_1998, lehur_entanglement_2008}. This viewpoint aligns with both numerical exact diagonalization studies and emerging quantum simulation platforms where only a finite number of bosonic modes is accessible~\cite{porras_mesoscopic_2008, sun_quantum_2025}. In these settings equilibration differs qualitatively from continuum baths since energy cannot dissipate irreversibly, and observables may exhibit persistent oscillations. However, the finite-dimensional many-body spectrum can be determined exactly, providing direct access to the energy-level structure that governs the system's quantum dynamics. Understanding how correlated equilibrium states emerge in these restricted environments is therefore a significant conceptual and practical challenge.

A key difficulty in this context is the preparation of the correlated ground state of the interacting system~\cite{abah_quantum_2020}. Starting from an uncoupled configuration and suddenly quenching the interaction leads to excitations that are difficult to disentangle from genuine dynamical responses. This can be problematic for studying controlled non-equilibrium processes, such as optical excitation or qubit manipulation~\cite{krantz_quantum_2019}. To address this, a common approach is the adiabatic switching protocol, in which the coupling between system and environment is gradually increased~\cite{rios_towards_2011, balzer_stopping_2016, karlsson_generalized_2018, tuovinen_adiabatic_2019, hopjan_initial_2019, joost_g1g2_2020, tuovinen_comparing_2020, tuovinen_electronic_2021, bonitz_accelerating_2024}. This allows the system to follow the instantaneous ground state, thereby preparing a correlated equilibrium state for further study. The success of this procedure is ultimately determined by the instantaneous many-body spectrum encountered during the switching process. In particular, regions where the low-energy spectrum exhibits avoided crossings are often accompanied by small excitation gaps, thereby imposing the most stringent conditions for adiabatic evolution.

\begin{figure}[t]
\includegraphics[width=\columnwidth]{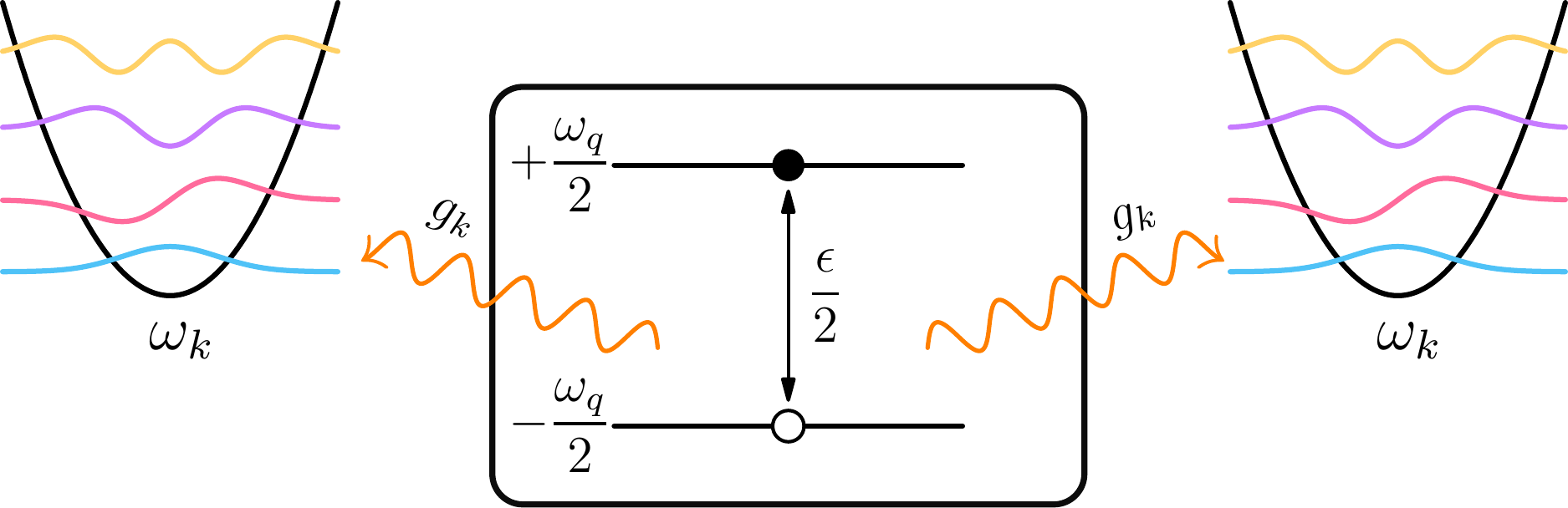}
\caption{Schematic of the spin-boson model: a two-level system coupled to a bath of harmonic oscillators.}
\label{fig:schematic}
\end{figure}

In this work, we investigate the efficiency and limitations of adiabatic preparation for correlated equilibrium states in finite spin-boson systems; see Fig.~\ref{fig:schematic}. We analyze the correlated many-body spectrum of the interacting system and relate its evolution with the spin-boson coupling strength to the conditions for adiabatic state preparation. We identify the spectral regions where the adiabatic constraints become most demanding and benchmark these predictions by analyzing the required switching times as a function of system parameters, including the spin-boson coupling strength and the bosonic mode structure.

The low-energy many-body spectrum governs the efficiency of adiabatic state preparation. In particular, avoided crossings lead to reduced excitation gaps and therefore characterize parameter regimes requiring significantly longer switching times. While the switching times extracted from the ground-state energy largely reflect the underlying spectral evolution, local observables such as Pauli matrix expectation values exhibit additional non-monotonic features associated with coherent oscillations in finite systems. These findings can serve as benchmarks for both theoretical treatments of open quantum systems and practical implementations in quantum simulation and qubit control.

\section{\label{sec:model and method}Spin-boson model and criteria for adiabatic state preparation}

\subsection{\label{subsec:model}Model Hamiltonian}

The Hamiltonian of the spin-boson model, depicted in Fig.~\ref{fig:schematic}, is given by
\begin{equation}\label{eq:Hamiltonian}
    \hat{H} = \frac{\omega_q}{2}\sigma_z + \frac{\epsilon}{2}\sigma_x + \sum_k\omega_k \hat{b}_k^\dagger \hat{b}_k + \sum_k g_k \sigma_x(\hat{b}_k + \hat{b}_k^\dagger),
\end{equation}
where $\sigma_z$ and $\sigma_x$ are Pauli matrices acting on the two-level system (qubit), $\omega_q$ is the energy splitting of the qubit, $\epsilon$ is the tunneling amplitude, $\omega_k$ are the bosonic mode energies, $\hat{b}_k^\dagger$ and $\hat{b}_k$ are the bosonic creation and annihilation operators of the bosonic (bath) modes, and $g_k$ is the coupling strength between the qubit and the $k$th bosonic mode.

The effect of the bath on the qubit is characterized by the spectral density
\begin{equation}
    J(\omega) = \sum_k |g_k|^2\delta(\omega - \omega_k).
\end{equation}
Here we consider the Lorentz-Drude spectral function \cite{breuer_theory_2002}
\begin{equation}\label{eq:spectral density}
    J(\omega) = \frac{\gamma \omega \omega_c^2}{\pi\omega_q(\omega^2 + \omega_c^2)},
\end{equation}
where $\gamma$ is the effective coupling strength and $\omega_c$ is the cut-off frequency. Since the effective coupling between the qubit and the bath is characterized by the spectral density, the coupling strengths $g_k$ are determined from it as a function of the effective coupling strength once the cut-off frequency and mode energies are fixed.

\subsection{\label{subsec:spectrum}Exact diagonalization: spectrum and avoided crossings}

In the absence of coupling, the system~\eqref{eq:Hamiltonian} is exactly solvable and the qubit and bath remain separable. Once the interaction is introduced, the qubit becomes dressed by bosonic excitations, giving rise to a correlated equilibrium ground state and rendering the problem nontrivial. This coupling-induced dressing can be conceptually related to polaron physics and can, in some regimes, be understood through the Lang-Firsov transformation as in electron-phonon problems~\cite{stefanucci_nonequilibrium_2013}. In general, however, the interacting spin-boson model does not admit an exact solution, although special cases such as the single-mode limit, corresponding to the Rabi model, are integrable~\cite{braak_integrability_2011}. For a sufficiently small number of bosonic modes, the system~\eqref{eq:Hamiltonian} can be diagonalized numerically, giving direct access to the exact ground state, which we later use as a benchmark for the time-dependent adiabatic preparation.

Throughout this work, we consider single and three bosonic modes, for which numerically exact diagonalization remains feasible. The qubit energy $\omega_q$ is used as the energy unit ($\omega_q=1$), such that time is measured in units of inverse qubit energy $\omega_q^{-1}$. The cut-off frequency in Eq.~\eqref{eq:spectral density} is fixed to $\omega_c=\omega_q$. The mode energies are set to $\omega_{k=1}=\omega_q$ in the single-mode case, and to $\omega_k=0.5k\,\omega_q$ with $k=1,2,3$ in the three-mode case.

The actual Hilbert space dimension of the spin-boson model must be truncated in practical calculations, because a single bosonic mode can be occupied by an arbitrary number of bosons. The effective Hilbert space is then given by the tensor product of the two-dimensional Hilbert space of the qubit and the $(N+1)^M$-dimensional Hilbert space of the truncated bath, where $N$ is the boson occupation per bosonic mode and $M$ the number of modes, giving a total dimension $\operatorname{dim}\mathcal{H} = 2(N+1)^M$. Physically, the truncation limits the maximum bosonic excitation dressing the qubit and must therefore be chosen sufficiently large to ensure convergence of the spectrum and of the relevant observables.

\begin{figure}[t]
\includegraphics[width=\columnwidth]{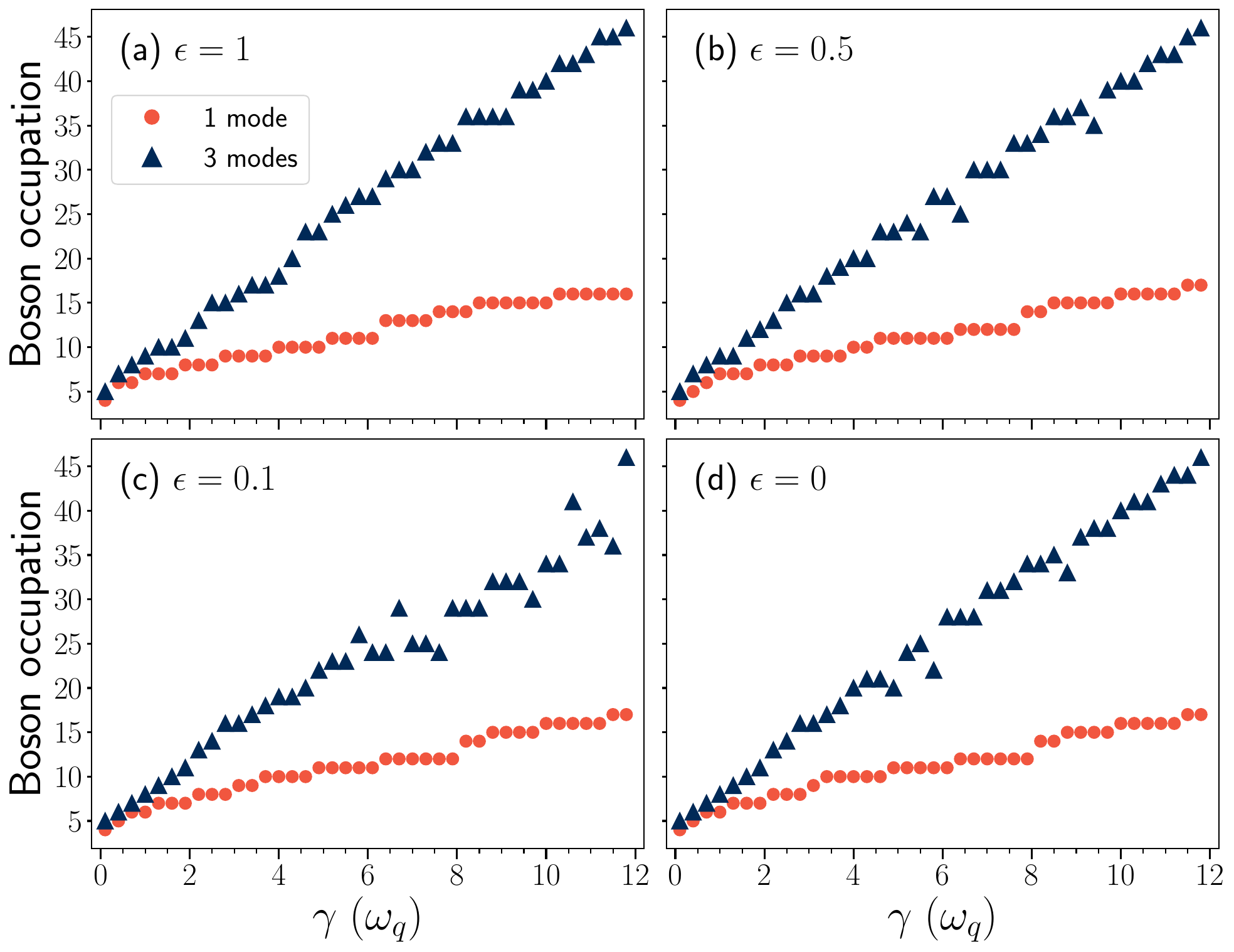}
\caption{\label{fig:boson_convergence}Convergence of the bosonic Hilbert space truncation as a function of the effective coupling strength $\gamma$ with different values of the tunneling amplitude $\epsilon$ for single and three bosonic modes.}
\end{figure}

Convergence of this truncation is verified in Fig.~\ref{fig:boson_convergence}. The cut-off is chosen such that the results are within a specified tolerance ($10^{-6}$), such that the calculations may be regarded as numerically exact within the chosen precision. The required basis size increases monotonically with the effective coupling strength, as stronger coupling regimes lead to more strongly dressed qubit-bath states and allow for higher bosonic occupations, thereby requiring a larger basis for convergence. In practice, this limits the accessible ranges of the effective coupling strength and the number of bosonic modes due to the increasing computational cost associated with the exponential growth of the effective Hilbert space. For example, for $\epsilon=1$, $\gamma=12$ and with three modes $M=3$, the required number of bosons per mode is $N=47$ (Fig.~\ref{fig:boson_convergence}(a)) and the total Hilbert space dimension is $\operatorname{dim}\mathcal{H}=221184$. Although larger systems could in principle be treated with increased computational resources, we restrict the calculations to parameter regimes for which convergence is achieved within the basis sizes shown in Fig.~\ref{fig:boson_convergence}, thereby maintaining controlled numerical accuracy at feasible computational cost. The energy spectra and adiabatic-condition matrix elements presented in the remainder of this section are computed using these converged basis sizes.

\begin{figure}[t]
\includegraphics[width=\columnwidth]{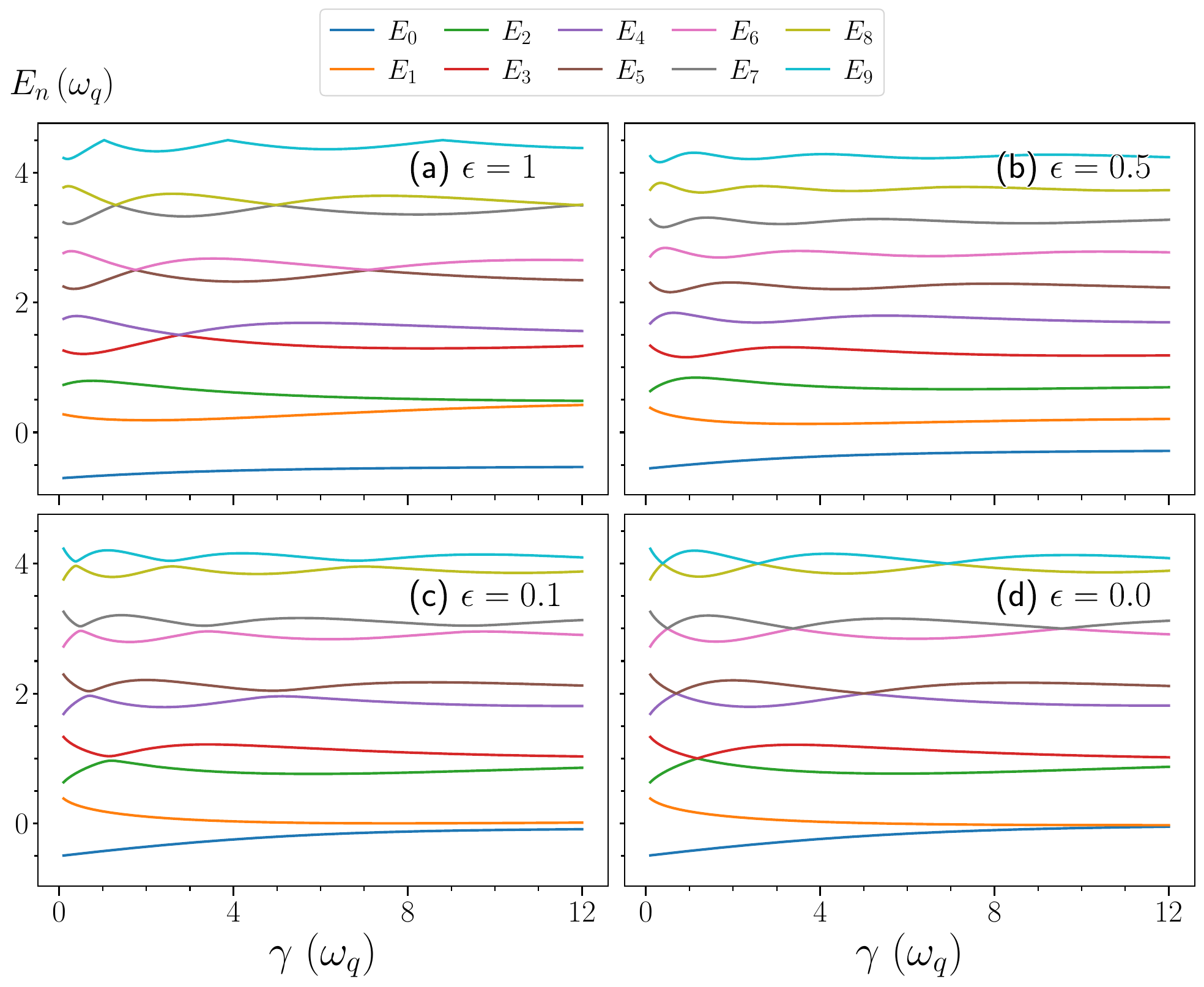}
\caption{\label{fig:exact diagonalization 1 mode}Energy spectra for a single bosonic mode, showing a series of avoided crossings depending on $\epsilon$ and $\gamma$. The spectrum is shifted by removing the linear contribution associated with the polaron shift $-\sum_k g_k^2 / \omega_k$.}
\end{figure}

We show the resulting single-mode spectrum in Fig.~\ref{fig:exact diagonalization 1 mode}, with the corresponding three-mode spectrum given in Fig.~\ref{fig:exact diagonalization 3 modes} in Appendix~\ref{sec:appendix A}. Already for a single bosonic mode, the spectrum exhibits a complex level structure with multiple avoided crossings, whose positions and gap sizes depend on the tunneling amplitude $\epsilon$ and the coupling strength $\gamma$.

\subsection{\label{subsec:adiabatic-condition}Adiabatic switching condition}

Alternatively, a ground state solution can be obtained through time-evolution by employing the adiabatic switching procedure: the system is initialized in the ground state of the non-interacting Hamiltonian $(g_k=0)$, after which the interaction is turned on slowly. If this is done slowly enough, the adiabatic theorem ensures that the system remains close to the instantaneous ground state and will eventually reach the correlated equilibrium ground state.

Time-dependence is introduced through the coupling term: replacing $g_k \to g_k(t)$ in the interaction part of Eq.~\eqref{eq:Hamiltonian} gives the time-dependent Hamiltonian
\begin{equation}
    \hat{H}(t) = \hat{H}_0 + \hat{H}_{\text{int}}(t), \quad \hat{H}_{\text{int}}(t) = \sum_k g_k(t) \sigma_x(\hat{b}_k + \hat{b}_k^\dagger),
\end{equation}
where $\hat{H}_0$ denotes the first three terms of Eq.~\eqref{eq:Hamiltonian}.

The coupling strength is ramped according to $g_k(t) = g_k f(t)$, with the ramp function $f:[0,t_0]\to[0,1]$ given by
\begin{equation}\label{eq:switch-on}
    f(t) =
    \begin{cases}
        \sin^2\left( \dfrac{\pi t}{2t_0} \right), & t \leq t_0 \\
        1, & t > t_0,
    \end{cases}
\end{equation}
where $t_0$ is the switch-on time. This connects the uncoupled Hamiltonian $\hat{H}_0$ at $t=0$ to the fully interacting Hamiltonian $\hat{H}_0+\hat{H}_{\text{int}}$, recovered at $t=t_0$ once $g_k(t)=g_k$. In the limit of infinitely slow switching, this construction is consistent with the Gell-Mann-Low theorem, which connects the noninteracting and interacting eigenstates through adiabatic continuation~\cite{gell-mann_bound_1951}. For finite switching times, however, the validity of the adiabatic approximation must be verified explicitly~\cite{stefanucci_nonequilibrium_2013}.

The requirement for sufficiently slow evolution can be quantified by the condition \cite{amin_consistency_2009, albash_adiabatic_2018}:
\begin{equation}\label{eq: adiabatic condition}
    \max_{s\in[0,1]}\frac{|\langle\psi_m(s)|\partial_s \hat{H}(s)|\psi_n(s)\rangle|}{|E_m(s) - E_n(s)|^2} \ll t_0, \quad m\neq n ,
\end{equation}
where $s=t/t_0$ is dimensionless time parameter, $\hat{H}(s)\equiv\hat{H}(t_0 s)$, and $|\psi_n(s)\rangle$ and $E_n(s)$ are the $n$-th instantaneous eigenstate and -energy of $\hat{H}(s)$,
\begin{equation}
    \hat{H}(s)|\psi_n(s)\rangle = E_n(s)|\psi_n(s)\rangle.
\end{equation}

\begin{figure}[t]
\includegraphics[width=\columnwidth]{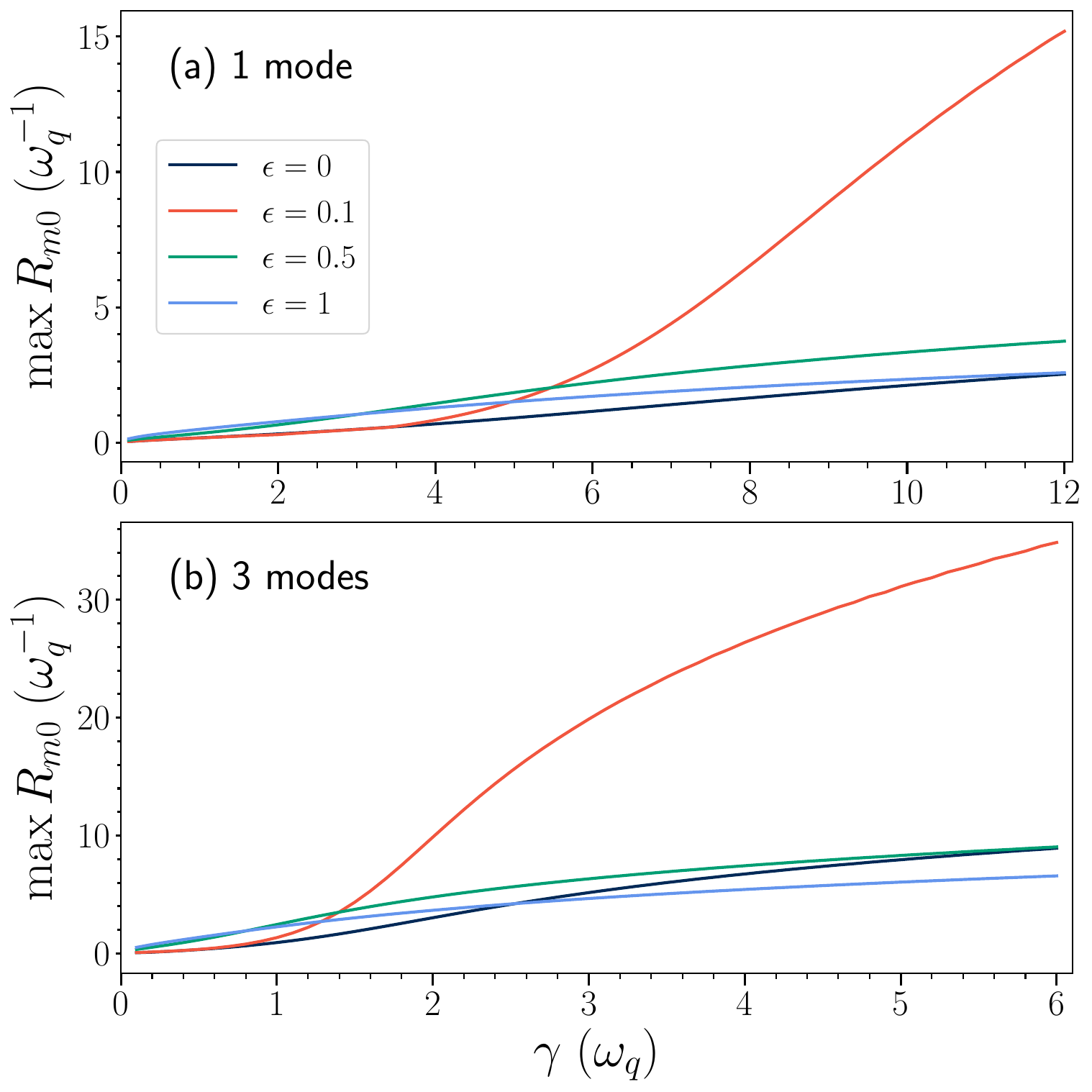}%
\caption{\label{fig:adiabatic condition}Maximum of $R_{m0}(s)$ on $s\in[0,1]$ as a function of the coupling $\gamma$ for $m=1,\dots ,9$ with (a) single mode and (b) three modes.}
\end{figure}

For the switching procedure in Eq.~\eqref{eq:switch-on}, the quantity to be maximized in the adiabatic condition, Eq.~\eqref{eq: adiabatic condition}, takes the form (see Appendix~\ref{sec:appendix A})
\begin{equation}\label{eq:r}
    R_{mn}(s) = \left|\frac{\pi}{2}\sin(\pi s)\right|\frac{|\langle\psi_m(s)|\hat{H}_{\text{int}}(s)|\psi_n(s)\rangle|}{|E_m(s) - E_n(s)|^2}.
\end{equation}
Figure~\ref{fig:adiabatic condition} shows $\max_{s\in [0,1]}R_{m0}(s)$ as a function of $\gamma$ with different values of the tunneling amplitude $\epsilon$ in the single- and three-mode cases. The results show similar behavior for different values of $\epsilon$, except for $\epsilon=0.1$, where the values become significantly larger when the coupling increases. Since the adiabatic condition~\eqref{eq: adiabatic condition} requires $t_0$ to exceed $\max_{s} R_{m0}(s)$, this signals that substantially longer switch-on times are needed to reach the ground state adiabatically in this parameter regime. We will return to this in Sec.~\ref{sec:results}, where the switch-on times are examined directly.

\subsection{\label{subsec:numerics}Numerical time evolution and convergence}

In principle, the time evolution of a quantum state vector $|\psi(t)\rangle$ is governed by the Schrödinger equation

\begin{equation}
    i\partial_t|\psi(t)\rangle = \hat{H}(t)|\psi(t)\rangle,
\end{equation}
with the formal solution
\begin{equation}
    |\psi(t_2)\rangle = \hat{U}(t_2,t_1)|\psi(t_1)\rangle
\end{equation}
where
\begin{equation}
    \hat{U}(t_2,t_1) = T\exp\left(-i\int_{t_1}^{t_2}d\bar{t}\,\hat{H}(\bar{t})\right)
\end{equation}
is the time-ordered propagator evolving the state from time $t_1$ to $t_2$. Equivalently, the dynamics can be obtained by solving the von Neumann equation
\begin{equation}\label{eq:von Neumann}
    \partial_t \hat{\rho}(t)
    =
    -i\left[\hat{H}(t),\hat{\rho}(t)\right]
\end{equation}
for the density operator $\hat{\rho}$ corresponding to a pure state $\hat\rho(t)=|\psi(t)\rangle\langle\psi(t)|$, extending to a statistical mixture $\hat{\rho}(t) = \sum_k p_k |\psi_k(t)\rangle\langle\psi_k(t)|$ provided every $|\psi_k(t)\rangle$ evolves under the same Hamiltonian $\hat{H}(t)$, i.e., for a closed system, as considered here~\footnote{In the general case, where environmental degrees of freedom are traced out, the reduced dynamics instead follows a master equation such as the Lindblad equation \cite{breuer_theory_2002}.}. 
The expectation value of an arbitrary operator $\hat{A}$ is then obtained as
\begin{equation}
    \langle \hat{A} \rangle(t)
    =
    \operatorname{Tr}\left[\hat{A}\hat{\rho}(t)\right].
\end{equation}
For numerical simulations, the operators are represented in a finite basis so that $\hat{H}$ and $\hat{\rho}$ become matrix representations of the corresponding operators. In practice, we solve the resulting equation of motion~\eqref{eq:von Neumann} numerically using the \texttt{mesolve} solver from the QuTiP library~\cite{qutip5, zenodo}.

Although we here consider a closed finite system, for which the von Neumann equation~\eqref{eq:von Neumann} is equivalent to the Schrödinger equation (as noted above), the density matrix formalism provides a more natural framework in the context of open quantum systems, where the bath degrees of freedom are typically integrated out, yielding a master equation such as the Lindblad equation for the reduced system density matrix \cite{breuer_theory_2002}.

\begin{figure}[t]
    \includegraphics[width=\columnwidth]{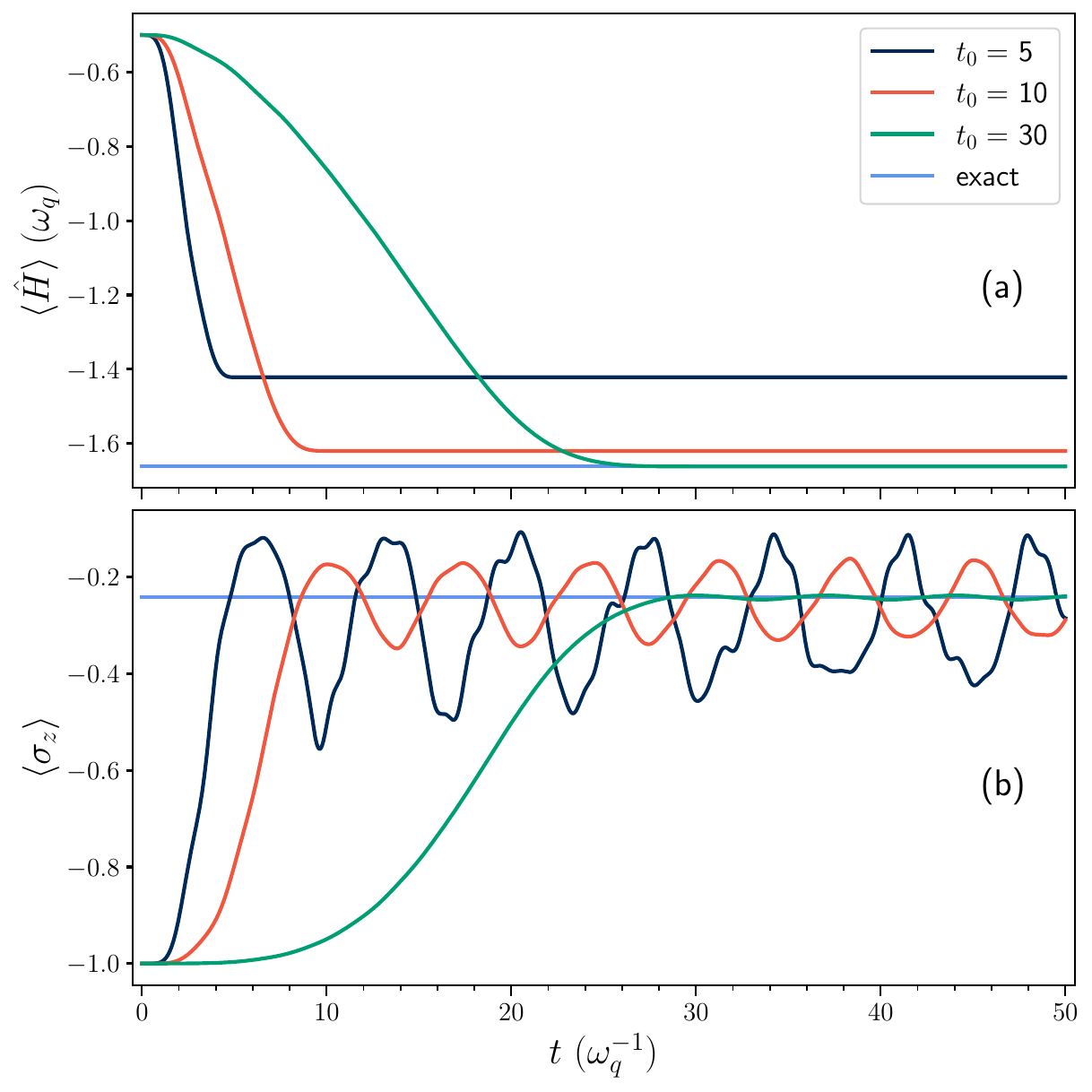}%
    \caption{Schematic illustration of the switch-on procedure. Time-dependent expectation values of (a) the total Hamiltonian and (b) the Pauli-Z matrix for various switching times $t_0$. Here, the evolutions are performed for a single bosonic mode with $\omega_q = \omega_k = \omega_c = \gamma = 1$ and $\epsilon=0$.}
    \label{fig:schematic procedure} 
\end{figure}

Figure~\ref{fig:schematic procedure} illustrates the switch-on procedure and the determination of the switch-on time. We first initialize the system and compute the exact ground state of the fully interacting Hamiltonian. Next, a switch-on time $t_0$ is chosen, and the system is time-evolved using the adiabatic switching procedure~\eqref{eq:switch-on} starting from the uncoupled state. The observable of interest is then evaluated using the evolved state and compared to the exact result. If the difference is within the specified tolerance, the result is accepted and convergence to the ground state is achieved; otherwise, $t_0$ is increased and the procedure is repeated until adiabatic convergence is reached. This operational criterion implements the notion of adiabatic state preparation, where sufficiently slow switching suppresses non-adiabatic transitions and ensures convergence to the instantaneous ground state~\cite{jansen_bounds_2007}.

In cases where the observable exhibits oscillatory behavior, the oscillation amplitude is defined as half the difference between its maximum and minimum values during the evolution. If the amplitude is below a specified tolerance, the time-averaged value is compared to the exact result. The oscillatory behavior and the role of the oscillation amplitude in determining the switch-on time are illustrated in Fig.~\ref{fig:schematic procedure}(b); compare, for example, the cases $t_0=10$ and $t_0=30$.

As noted above, we focus on single and three bosonic modes, for which the exact ground state is numerically accessible via diagonalization and serves as a reference for the adiabatically prepared state. For larger baths, exact diagonalization becomes infeasible, so no such reference is available for comparison.

\section{\label{sec:results}Consequences of avoided crossings for switching times}

\begin{figure*}[t]
\includegraphics[width=\textwidth]{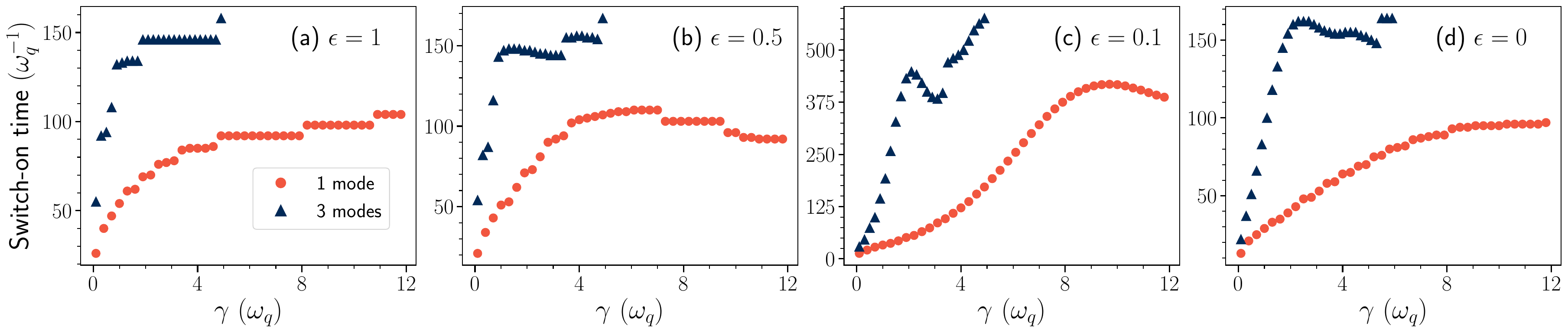}%
\caption{\label{fig:switch times gs}
Switch-on times as a function of the effective coupling strength $\gamma$ for single and three bosonic modes. The switch-on times are extracted using the ground state energy as a criterion for saturation. 
(a) $\epsilon=1$, (b) $\epsilon=0.5$, (c) $\epsilon=0.1$ and (d) $\epsilon=0$.}
\end{figure*}

Using the basis sizes required for convergence (Fig.~\ref{fig:boson_convergence}), we calculate the switch-on times using the adiabatic procedure illustrated in Fig.~\ref{fig:schematic procedure}(a) with a tolerance of $10^{-6}$. The resulting switch-on time provides a measure of the timescale required to suppress non-adiabatic excitations and achieve convergence to the correlated equilibrium state within the specified accuracy. Figure~\ref{fig:switch times gs} shows the required switch-on times as a function of the effective coupling strength $\gamma$ for single and three bosonic modes and for different values of the tunneling amplitude $\epsilon$.

In the case $\epsilon=1$ (Fig.~\ref{fig:switch times gs}(a)), where the tunneling amplitude is equal to the qubit energy $\omega_q$, the switch-on time increases monotonically with the effective coupling strength. This trend is consistent with the stronger qubit-bath hybridization at larger coupling strengths, which generally requires slower switching to suppress non-adiabatic excitations, and mirrors the monotonic growth of $R_{m0}(s)$ with $\gamma$ in Fig.~\ref{fig:adiabatic condition}(a). The three-mode case requires nearly twice the switch-on time compared to the single-mode case. This can be associated with the denser set of avoided crossings in the three-mode spectrum compared to the single-mode case (Fig.~\ref{fig:exact diagonalization 3 modes}(a)). Near an avoided crossing the energy gap $|E_m(s)-E_n(s)|$ is small while the coupling matrix element $|\langle\psi_m(s)|H_{\text{int}}|\psi_n(s)\rangle|$ remains finite, so $R_{mn}(s)$ grows sharply, as seen in Fig.~\ref{fig:adiabatic condition}. Through the adiabatic condition, Eq.~\eqref{eq: adiabatic condition}, this directly translates into a longer required switch-on time $t_0$.

In the absence of the tunneling term, $\epsilon=0$ (Fig.~\ref{fig:switch times gs}(d)), the behavior is similar for the single-mode case, whereas for the three-mode case the switch-on time increases monotonically for $\gamma\leq 2.5$, while exhibiting non-monotonic behavior in the $\gamma=2-6$ range. This coincides with the energy-level structure shown in Fig.~\ref{fig:exact diagonalization 3 modes}(d): in this range, avoided crossings and a bunching of energy levels reduce the spacing between excited states as $\gamma$ increases. Since $R_{mn}(s)$ scales inversely with $|E_m(s)-E_n(s)|^2$, such level bunching can locally increase the adiabatic condition, Eq.~\eqref{eq: adiabatic condition}, even though $R_{m0}(s)$ itself increases monotonically with $\gamma$ in Fig.~\ref{fig:adiabatic condition}(b), since the bound is set by the excited state that maximizes the corresponding adiabatic ratio. In this case, the smallest ground-state gap does not directly determine the bound: the matrix element $|\langle\psi_1(s)|H_{\mathrm{int}}|\psi_0(s)\rangle|$ is numerically negligible, so the lowest excited state contributes little to $R_{m0}(s)$ and the bound is instead determined by higher excited states.

When the tunneling amplitude is half of the qubit energy, $\epsilon=0.5$ (Fig.~\ref{fig:switch times gs}(b)), the behavior is monotonic for $\gamma\leq 6$ and for $\gamma\leq 1.3$ in the single- and three-mode cases, respectively. For larger coupling strengths, the switch-on time decreases slightly before increasing again. Similar behavior is observed when the tunneling amplitude is an order of magnitude smaller than the qubit energy, $\epsilon=0.1$ (Fig.~\ref{fig:switch times gs}(c)), where monotonic increase persists up to $\gamma\leq 9.6$ and $\gamma\leq 2.1$ in the single- and three-mode cases, respectively. As anticipated in Sec.~\ref{subsec:adiabatic-condition}, $R_{m0}(s)$ in Fig.~\ref{fig:adiabatic condition} is largest for $\epsilon=0.1$, consistent with the growth in the switch-on time up to the largest couplings shown here. Unlike the $\epsilon=1$ case, where the switch-on time increases monotonically throughout, both $\epsilon=0.5$ and $\epsilon=0.1$ exhibit local reductions in the switch-on time, which shift to larger coupling strengths as $\epsilon$ decreases. As in the $\epsilon=0$ case, these local reductions coincide with avoided crossings and level bunching in the corresponding spectra (Fig.~\ref{fig:exact diagonalization 1 mode}(b,c) and Fig.~\ref{fig:exact diagonalization 3 modes}(b,c)). For $\epsilon=0.1$, the excited state providing the largest contribution to the adiabatic condition also changes with $\gamma$: the maximum is associated with $m=2$ for $\gamma\lesssim3.5$, but with the first excited state for larger couplings. The latter coincides with the narrowing of the $E_0$-$E_1$ gap for $\gamma \gtrsim 4$ visible in Fig.~\ref{fig:exact diagonalization 1 mode}(c). In the three-mode case this effect is slightly more pronounced, consistent with the additional freedom in distributing bosonic excitations across modes and with the denser level structure already noted for $\epsilon=1$.

The switch-on times are on similar time scales except for $\epsilon=0.1$, for which they are approximately three times larger. Compared to the $\epsilon=0$ case, already a small but nonzero tunneling amplitude induces mixing of the qubit states, which modifies the dynamical pathways through which the system approaches the interacting ground state. In this regime, we have checked (not shown) that the evolution remains smooth and largely free of oscillatory behavior, but convergence to the exact ground-state expectation values requires significantly longer switch-on times. This indicates that adiabaticity is limited not by strong non-adiabatic oscillations, but rather by a slow approach to saturation of the relevant observables, leading to an effective separation of time scales in the adiabatic preparation. In contrast to the $\epsilon=1$ and $\epsilon=0$ cases, this intermediate regime therefore requires substantially longer evolution times to reach the same level of convergence. This difference in switch-on times clearly connects with $\max_{s\in[0,1]} R_{m0}(s)$ in Fig.~\ref{fig:adiabatic condition}, as anticipated in Sec.~\ref{subsec:adiabatic-condition}. While comparable in magnitude for small $\gamma$, the values grow faster for $\epsilon=0.1$. Through the adiabatic condition~\eqref{eq: adiabatic condition} this larger bound is consistent with the observed longer switch-on times.

\begin{figure}[b]
\includegraphics[width=\columnwidth]{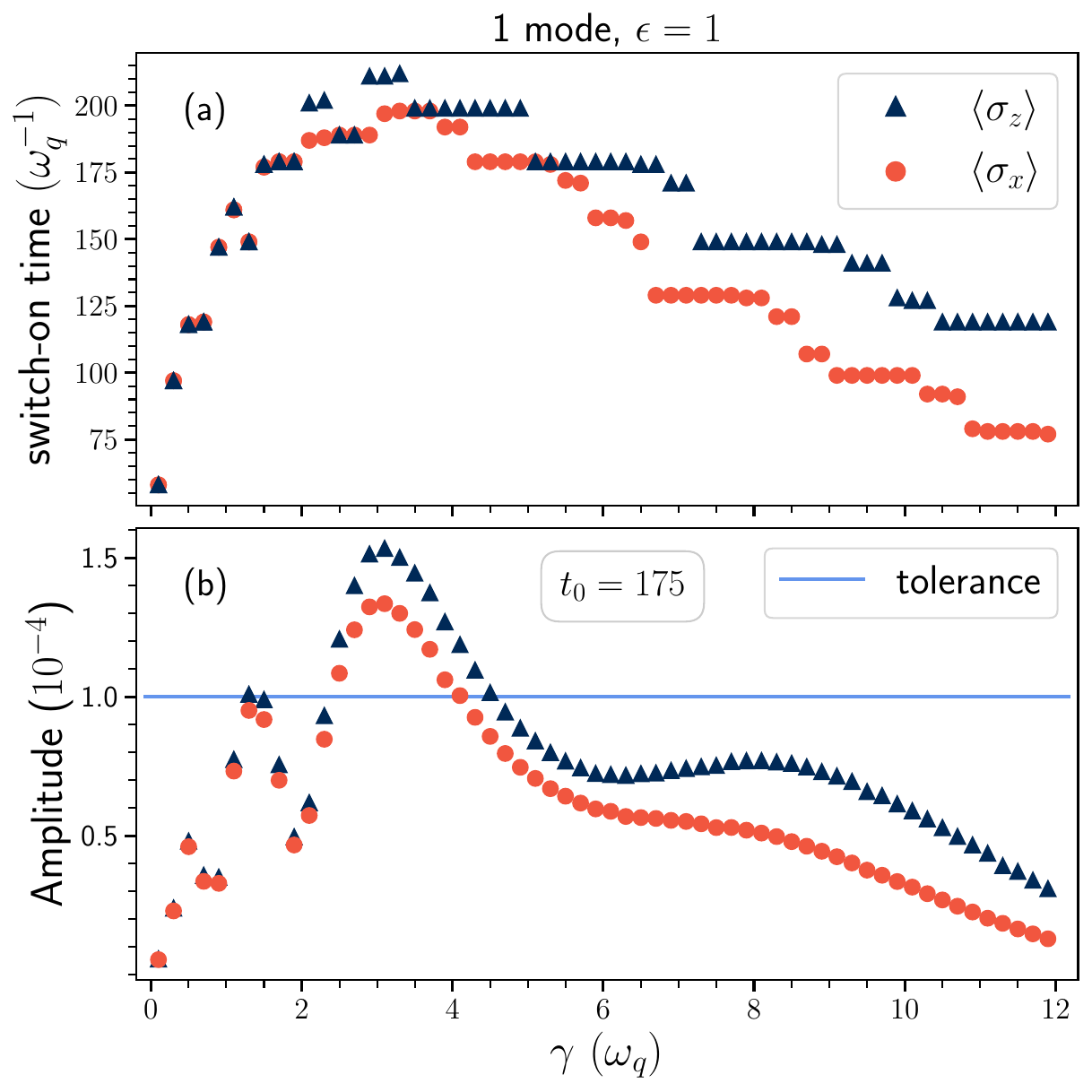}%
\caption{\label{fig:switch times sigmaz}(a) The switch-on times as a function of the effective coupling strength $\gamma$ for single mode. The switch-on times are extracted using $\langle\sigma_z\rangle$ and $\langle\sigma_z\rangle$ as a criterion for saturation. (b) Amplitudes of the oscillations of $\langle\sigma_z\rangle$ and $\langle\sigma_x\rangle$ with switch-on time $t_0=175$ and tolerance $10^{-4}$.}
\end{figure}

The previous results were obtained using the ground-state energy as the target observable. Figure~\ref{fig:switch times sigmaz}(a) instead shows the switch-on times extracted using the expectation values of the Pauli matrices. The convergence criteria were chosen as $10^{-6}$ for the difference between the time-averaged value and the exact results and $10^{-4}$ for the oscillation amplitude. We have verified (not shown) that varying the amplitude tolerance does not qualitatively affect the results in terms of the effective coupling strength $\gamma$, it only rigidly shifts the required switch-on time. The corresponding adiabatic procedure is illustrated in Fig. ~\ref{fig:schematic procedure}(b).

We calculated the switch-on times for $\epsilon=1$ in the single-mode case using the expectation values of $\sigma_z$ and $\sigma_x$. In contrast to the results obtained using the ground-state energy, the switch-on time depends non-monotonically on $\gamma$: it increases up to a critical value, beyond which it decreases again. The critical value $\gamma\sim3.3$ corresponds to a coupling strength of $g_k\sim \omega_q$, where the qubit-bath coupling becomes comparable to the intrinsic energy scale of the qubit (cf.~Eq.~\eqref{eq:spectral density}), indicating a near-resonant regime between the qubit splitting and the bath mode energy. 

The critical value lies between the avoided crossings in the single-mode spectrum at $\epsilon=1$ ($E_3$-$E_4$ at $\gamma\approx2.7$ and $E_7$-$E_8$ at $\gamma\approx5$; Fig.~\ref{fig:exact diagonalization 1 mode}(a)), where excited-state gaps remain comparatively narrow. Coherences between such near-degenerate states oscillate at a frequency set by the local gap, and can therefore sustain persistent oscillations in this range, consistent with the amplitude criterion set in Sec.~\ref{subsec:numerics}: a narrower gap requires a longer switch-on time to average below the $10^{-4}$ tolerance, even though the ground-state energy converges smoothly over the same range of $\gamma$. This is verified in Fig.~\ref{fig:switch times sigmaz}(b) for $t_0=175$, where the oscillation amplitudes near the critical value exceed the tolerance threshold $10^{-4}$, leading to the requirement of longer switch-on times in order to satisfy the convergence condition, as shown in Fig.~\ref{fig:switch times sigmaz}(a).

\section{\label{sec:conculsions}Conclusions}

We examined the performance of adiabatic switching protocols for preparing correlated equilibrium states in finite spin-boson systems by comparing the resulting dynamics against exact diagonalization benchmarks over a range of coupling strengths and bath mode configurations. The energy spectrum of the spin-boson model exhibits a series of avoided crossings, whose positions and gap sizes depend on the tunneling amplitude and coupling strength. Through the adiabatic condition, these narrowing gaps directly govern the timescale required for adiabatic state preparation. A central outcome is that the preparation of global quantities follows expectations from conventional adiabatic theory: the timescale required to accurately recover the ground-state energy grows steadily as the spin-boson interaction becomes stronger, consistent with the denser avoided-crossing structure and narrower gaps found at stronger coupling and weaker tunneling, reflecting the increasing difficulty of adiabatically reaching strongly correlated regimes.

We also found that the apparent efficiency of state preparation may depend on the observable used to assess convergence. In contrast to the monotonic behavior of the total energy, local quantities such as Pauli matrix expectation values may display non-monotonic switching-time behavior. These deviations arise from coherent oscillatory dynamics sustained near the narrow gaps associated with avoided crossings and finite-size effects associated with the discrete bosonic bath, demonstrating that local observables can equilibrate on substantially different timescales than global energetic measures.

This analysis provides a starting point for exploring the genuine nonequilibrium dynamics of qubit systems prepared in correlated equilibrium states. In practical quantum-computing platforms, qubits are often manipulated through externally applied optical or thermal excitations and subsequently probed through related readout schemes~\cite{gao_coherent_2015, gritsch_optical_2025, arnold_all-optical_2025, wang_longitudinal_2025, gunyho_single-shot_2024, aamir_thermally_2025}. Starting from a correlated equilibrium state, rather than an initially uncoupled configuration, offers a controlled route for investigating the response of the system to external driving while avoiding transient effects associated with artificial initial conditions. The dynamical evolution can thus be more directly linked to the interplay between the external perturbation and the intrinsic quantum properties of the coupled spin-boson system, paving the way for more accurate protocols in qubit control and quantum information processing.

\begin{acknowledgments}
We acknowledge the financial support of the Finnish Ministry of Education and Culture through the Quantum Doctoral Education Pilot Program (QDOC VN/3137/2024-OKM-4) and the Research Council of Finland through the Finnish Quantum Flagship project (359240, JYU). R.v.L. acknowledges support from the Research Council of Finland under Project No. 356906. R. T. acknowledges support from the Jane and Aatos Erkko Foundation under Project EffQSim. We acknowledge grants of computer capacity from the Finnish Grid and Cloud Infrastructure (persistent identifier urn:nbn:fi:research-infras-2016072533).
\end{acknowledgments}

The data that support the findings of this article are openly available~\cite{zenodo}.

\appendix

\section{\label{sec:appendix A}Switching protocol conditions and auxiliary results}

Starting from the adiabatic condition~\eqref{eq: adiabatic condition}, we define the quantity to be maximized, $R_{mn}(s),$ as
\begin{equation}
    R_{mn}(s) = \frac{|\langle\psi_m(s)|\partial_s \hat{H}(s)|\psi_n(s)\rangle|}{|E_m(s) - E_n(s)|^2}.
\end{equation}
In terms of the dimensionless time parameter $s=t/t_0$, the switching function in Eq.~\eqref{eq:switch-on} becomes
\begin{equation}
    f(s) = \sin^2\left( \frac{\pi s}{2} \right).
\end{equation}
Consequently,
\begin{equation}
    \partial_s \hat{H}(s) = \frac{\pi}{2}\sin(\pi s) \hat{H}_{\text{int}}
\end{equation}
and
\begin{equation}\label{eq: R_mn}
    R_{mn}(s) = \left|\frac{\pi}{2}\sin(\pi s)\right|\frac{|\langle\psi_m(s)|\hat{H}_{\text{int}}|\psi_n(s)\rangle|}{|E_m(s) - E_n(s)|^2}.
\end{equation}

\begin{figure}
\includegraphics[width=\columnwidth]{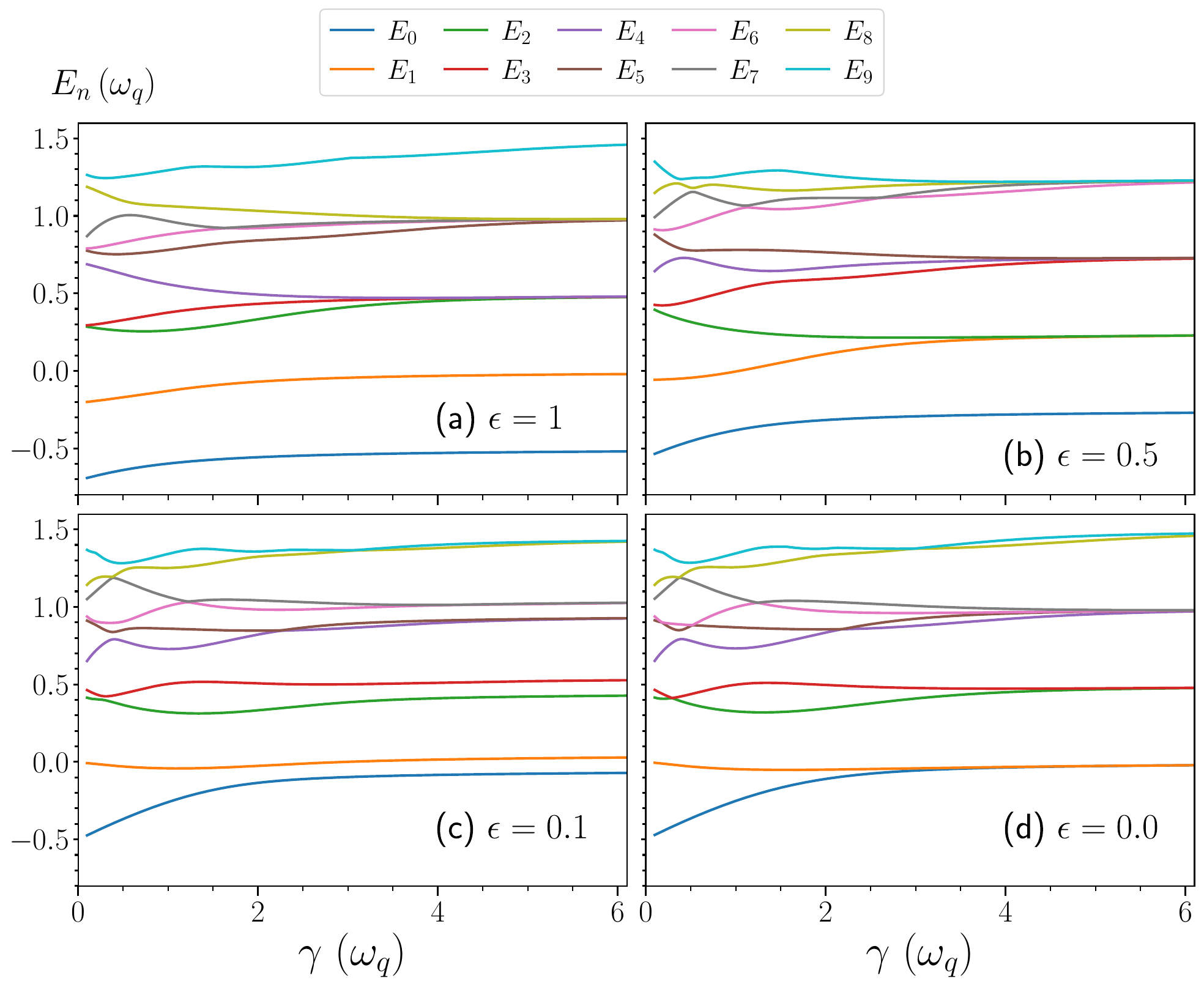}%
\caption{\label{fig:exact diagonalization 3 modes}Energy spectra for three bosonic modes. Obtained similarly as Fig.~\ref{fig:exact diagonalization 1 mode}.}
\end{figure}

Similar to Fig.~\ref{fig:exact diagonalization 1 mode} in the main text, we show the spectra for three bosonic modes in Fig.~\ref{fig:exact diagonalization 3 modes}.

\end{document}